\documentclass{article}
\usepackage[letterpaper, margin=1in]{geometry}

\usepackage{cite}
\usepackage{amsmath,amssymb,amsfonts}
\usepackage{algorithmic}
\usepackage{graphicx}
\usepackage{textcomp}
\usepackage{xcolor}
\def\BibTeX{{\rm B\kern-.05em{\sc i\kern-.025em b}\kern-.08em
    T\kern-.1667em\lower.7ex\hbox{E}\kern-.125emX}}
\begin{document}

\title{Anomaly Detection in Cloud Components}

\author{Mohammad Saiful Islam and Andriy Miranskyy \\
Department of Computer Science, Ryerson University, Toronto, Canada \\
mohammad.s.islam@ryerson.ca, avm@ryerson.ca}

\date{}

\maketitle

\begin{abstract}
Cloud platforms, under the hood, consist of a complex inter-connected stack of hardware and software components. Each of these components can fail which may lead to an outage. Our goal is to improve the quality of Cloud services through early detection of such failures by analyzing resource utilization metrics. We tested Gated-Recurrent-Unit-based autoencoder with a likelihood function to detect anomalies in various multi-dimensional time series and achieved high performance.
\end{abstract}

\section{Introduction}
Digitalization has changed the way we have looked at things for centuries. Computers, electronic gadgets, the Internet, IoT devices are part and parcel of our modern lifestyle. Cloud computing systems play a significant role in meeting the rapidly growing technological demand. Such services are offered by remote servers and shared by customers on the Internet \cite{mell2011nist}. Cloud services are becoming increasingly popular as government and businesses move their facilities to Cloud platforms. The market size of public Cloud services is worth approximately \$160 billion globally in 2018, and is predicted to reach \$277 billion by 2021 at an annual growth rate of 21.9\% \cite{idc2018worldwide}. Therefore, these platforms are becoming critical to the operations of entities using them.

\subsection{Big Data Challenge}
In recent years, the advanced monitoring system has been deployed in data centers to observe the health of Cloud components. However,
most current health monitoring systems still rely on statistics and heuristics based on resource utilization thresholds only from the observed health metrics \cite{pourmajidi2017challenges, pourmajidi2019dogfooding}. The variety and the
random nature of today's application requirements make such anomaly detection techniques less effective in a Cloud environment. Moreover, the growing complexity, scale, speed, and volume of log and metrics data generated by components of Cloud platforms makes an analysis of such data a Big Data challenge \cite{miranskyy2016operational}.

\subsection{Logs and Metrics}
Logs usually refer to a collection of system-generated sets of data to describe an event that happened \cite{chuvakin2012logging}. Chuvakin et al. denotes \cite{chuvakin2012logging} the general structure of a log message: Timestamp, Source, and Data. The event details might include resource utilization information, the user who accessed that, and other application related facts.

Metrics are measurement at a point in time for the behaviours and conditions of a particular system. In place of continually gathering all the metadata about the entire system's health, a parameter will normalize this to get an only metric. Therefore, in place of logging the whole data over and over, metrics will only store the number and timestamp at regular intervals. Average response time per minute, hourly memory, and CPU utilization of a server, incoming traffic every ten seconds in a web server are some example metrics.

\subsection{Outlier Detection}
Anomalies, also known as outliers, are unusual patterns in a dataset that do not conform to the expected behaviour and one of the oldest statistical problems \cite{chandola2009anomaly}.  The outliers in the system could occur due to several reasons \cite{nousiainen2009anomaly} including software failures, hardware malfunctioning, system overload, human error, and noise.

Anomaly detection is part of many data mining techniques, and it can be used in businesses, such as intrusion detection in network traffic, health monitoring, pattern recognition in diagnosis, and fraud detection in financial transactions \cite{chandola2009anomaly}. Efforts have been made to detect anomalies in time series using statistical and neural network approaches. Most of the neural network models deal with single-dimensional time series. For Cloud components, one often needs to correlate multiple metrics (e.g., CPU, memory, and network utilization), hence the need for a model capable of processing multidimensional time series.

Our \textbf{goal} is to find anomalies in this multidimensional Cloud resource utilization time-series data using Machine Learning tools and techniques. We have achieved that by creating a dimension-independent neural network model to detect outliers in components of a Cloud platform.

\section{Literature Review}
Time series is generally a series of time-indexed data points consisting of a sequence of observations taken at equally spaced successive time intervals \cite{box2015time}. They are studied in different disciplines, such as science, engineering, and economics. Applied science and engineering fields like communication, signal processing, weather forecasting and earthquake prediction rely deeply on time series data to find out patterns and trends.

Massive amounts of logs and metrics are generated from modern systems. It is not feasible to get the data points labelled by experts, considering the scale. So, unsupervised learning methods seem rational to auto label the logs and raise flags to operation teams for potential anomalies. Farshchi et al.~\cite{farshchi2018metric} review the complexity of modern large-scale applications on the Cloud and indicate that traditional system monitoring approaches suffer from many limitations in complex Cloud environments, as handling non-stationary multidimensional time series by classic tools, such as a Seasonal Auto-Regressive Integrated Moving Average (SARIMA), is a non-trivial task. Instead, they may be better handled by neural networks~\cite{wang2011solar}. 

The works that are closest to ours are as follows. Numenta Platform for Intelligent Computing utilized Hierarchical Temporal Memory (HTM) \cite{ahmad2017unsupervised} based learning algorithm for error calculation in time series. They have implemented a likelihood function that processes raw error incrementally and uses historical errors modelled as a normal distribution. Kyle et al. \cite{hundman2018detecting} from NASA demonstrates the effectiveness of Long Short-Term Memory (LSTMs) networks-based anomaly detectors along with the Numenta Anomaly Benchmark (NAB) Likelihood function \cite{ahmad2017unsupervised}. They have used expert-labelled telemetry anomaly data from the Soil Moisture Active Passive satellite and the Mars Science Laboratory rover, Curiosity.

However, these papers are complementary to ours, because HTM processes one-dimensional data, while NASA's multi-dimensional LSTM models are tested in a domain different from ours (i.e., ``rocket science'' vs. ``Cloud components'').

\section{Methodology}
We are building a scalable machine-learning-based anomaly detector that can automatically detect failures of Cloud components. We have examined standard statistical and machine learning algorithms, e.g., SARIMA, artificial neural network \cite{hamzaccebi2008improving}, LSTM, and Gated Recurrent Units (GRU) \cite{cho2014learning} models to analyze time-series data and identify anomalies. Eventually, we focused on GRU-based models as, based on our experience, they gave the strongest predictive power out of all the models that we experimented with. We have explored the parameters controlling GRU architecture that yield the best predictive power for different time series. Finally, we look for an adequate window size of input data: that is how many observations in the time series are needed to capture complex patterns in time series while not slowing the computations significantly.

\subsection{Evaluation Metric}
We use NAB scoring \cite{ahmad2017unsupervised} to evaluate the performance of anomaly detectors. Ahmad et al. explained \cite{ahmad2017unsupervised} that classic metrics, e.g., Precision and Recall, are not suitable for the assessment of anomaly
detectors as they do not reward early detection.

\subsection{Anomaly likelihood}
We use anomaly likelihood measurement approach introduced by~\cite{ahmad2017unsupervised}. It maintains a window of the last $W$ error values and processes raw errors incrementally. Historical errors are modelled as a rolling normal distribution of a window of the last $W$ points at each step $t$. The empirical mean $\mu_t$ and standard deviation $\sigma_t$ at time $t$ are computed as follows:
\begin{equation}\label{eq:mu}
    \mu_t = \frac{\sum_{i=0}^{W-1}{s_{t-i}}}{W},
\end{equation}
where $s_{(\cdot)}$ is the prediction error computed by the model, and
\begin{equation}\label{eq:sigma}
    \sigma_t =  \sqrt{\frac{\sum_{i=0}^{W-1}{({s_{t-i} - \mu_t})^2}}{W-1}}.
\end{equation}
Akin to Eq.~\ref{eq:mu}, we compute empirical mean for a moving window $W^{'}$, deemed $\Tilde{\mu}_t$ . By design $W^{'} \ll W$; i.e., $W$ and $W^{'}$ are long- and short-term intervals, respectively. 

The likelihood of anomaly at time $t$, deemed $L_t$, is  
\begin{equation}\label{eq:Likelihood}
    L_t = 1 - Q\left(\frac{\Tilde{\mu}_t - \mu_t}{{\sigma}_t}\right), L_t\in(0,1),
\end{equation}
where  $Q$ is a Gaussian tail probability \cite{karagiannidis2007improved}. 
For a  user-defined threshold $\epsilon$, if $L_t \geq 1 - \epsilon$, an observation at time $t$ is classified as anomalous.

\begin{figure*}[t]
\centerline{\includegraphics[width=1.0\textwidth]{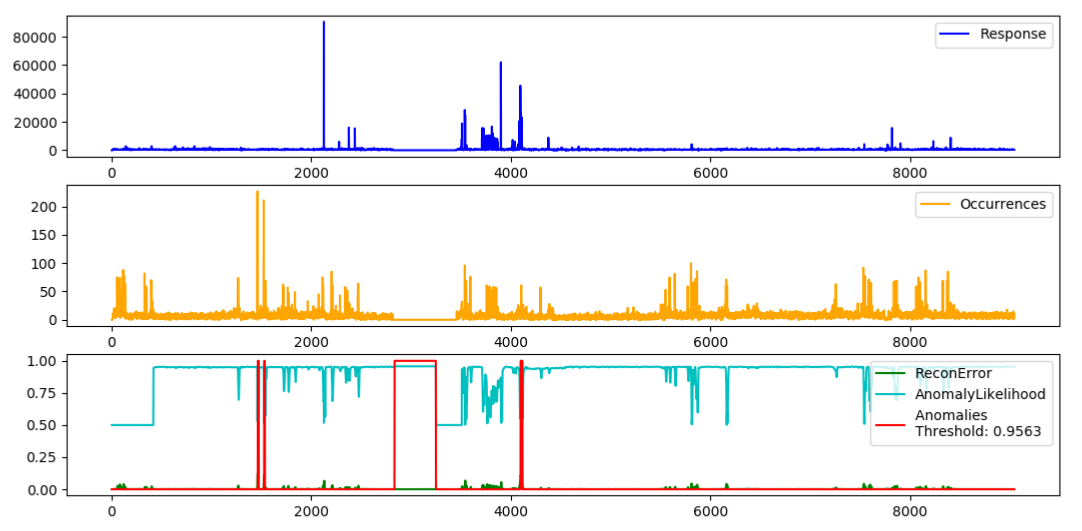}}
\caption{Anomaly detection in Web server response time of private Cloud. Top graph (in blue) indicates response time in milliseconds where the middle one (in orange)
shows frequencies. The bottom graph shows reconstruction error (in green), likelihood score (in cyan) and anomalies with the likelihood threshold of
0.9563 (in red). Timestamps marked with an index on the $x$-axis for better readability for all three graphs.}
\label{fig}
\end{figure*}

\subsection{Dataset}
\subsubsection{One-dimensional dataset} We used a popular reference NAB dataset \cite{numenta2017nab}, composed of 58 labelled real-world and artificial one-dimensional time-series data files that can be used for the evaluation of algorithms for anomaly detection. This data file contains two columns: timestamp and value. Data has been captured in 5 minutes intervals.
\subsubsection{Multi-dimensional dataset} We have collected these data from components of a private Cloud. We gathered one month's worth of data. The components are virtual machines running SaaS software. The metrics gathered include but are not limited to CPU utilization, memory consumption, and the response time of the SaaS application. We aggregated the data in 5-minute intervals (based on the suggestion of the Ops team running the software). The Ops team also provided us with feedback on the anomalies in the time series.

\subsection{Data Cleaning and Preprocessing}
One-dimensional NAB data was clean without missing values. We have marked the response time of multidimensional private Cloud data as minus one for the absence of any request in any given ten minutes interval.  We have normalized the data using min-max normalization before training the models.

\subsection{Results}
We used the GRU-based auto-encoder to get the reconstruction error; then, we have marked the anomalies based on the error values. We saw that the predictive power of the raw error values is low (NAB score $< 40$). Therefore, a transformation of these values was in order. We calculate the likelihood of anomaly by analyzing the distribution of the reconstruction error scores generated by the autoencoder as per \cite{ahmad2017unsupervised}.

\subsubsection{One-dimensional dataset} We achieved NAB score of 59.8, a third-best score for the one-dimensional dataset (as per NAB dashboard~\cite{numenta2017nab}). This shows that the model yields adequate results for the one-dimensional case.
\subsubsection{Multi-dimensional dataset} To preserve space, we show an example of our GRU-based model on two dimensions of the multidimensional time series (namely, average response time and frequency of requests in a 5-minute interval) in Figure~\ref{fig}. Our model can detect anomalies for approximately 70\% of the cases, which practitioners found satisfactory (in comparison
with the existing approaches that they used).

\section{Conclusion}
We created a GRU-based model for anomaly detection in Cloud components based on the analysis of multi-dimensional telemetry. The performance of the models in our experiments was evaluated on publicly available benchmark and the data gathered from private Cloud providers, yielding strong results for one- and multi-dimensional data.

Our work is of interest to practitioners, as better anomaly detectors of Cloud components will improve platform monitoring, leading to reduced number of outages, hence the improved customer satisfaction and increased competitiveness. Our work is also of interest to academics, as it serves as a building block in the theory of maintaining complex Cloud solutions.

\bibliographystyle{IEEEtran}
\bibliography{references} 

\end{document}